%% file: dpf2013_proceedings.tex
\documentclass[12pt]{article}
\pdfoutput=1

\usepackage{graphicx}
\usepackage{cite}

\newcommand\pubnumber{DPF2013-255}
\newcommand\pubdate{October 1, 2013}

\def\scipp{Santa Cruz Institute for Particle Physics \\
		   University of California, Santa Cruz}
\def\behalf{\footnote{On behalf of the Heavy Photon Search Collaboration}}

\def\Title#1{\begin{center} {\Large #1 } \end{center}}
\def\Author#1{\begin{center}{ \sc #1} \end{center}}
\def\Address#1{\begin{center}{ \it #1} \end{center}}

\newcommand\pubblock{\rightline{\begin{tabular}{l} \pubnumber\\
         \pubdate  \end{tabular}}}
\newenvironment{Abstract}{\begin{quotation}  }{\end{quotation}}
\newenvironment{Presented}{\begin{quotation} \begin{center} 
             PRESENTED AT\end{center}\bigskip 
      \begin{center}\begin{large}}{\end{large}\end{center} \end{quotation}}

\input econfmacros.tex

\begin{document}

\begin{titlepage}
\pubblock

\vfill
\Title{The Heavy Photon Search Experiment at Jefferson Lab}
\vfill
\Author{Omar Moreno\behalf}
\Address{\scipp}
\vfill
\begin{Abstract}
The Heavy Photon Search (HPS) is a new experiment at Jefferson Lab that will
search for heavy $U(1)$ vector bosons (heavy photons or dark photons) in the
mass range of 20 MeV/$c^2$ to 1 GeV/$c^2$. Dark photons in this mass range are 
theoretically favorable and may mediate dark matter interactions. The dark 
photon couples to electric charge through kinetic mixing with the photon, 
allowing its production through a process analogous to bremsstrahlung 
radiation. HPS will utilize this production mechanism to probe dark photons
with relative couplings of $\varepsilon^2 = \alpha'$/$\alpha \sim$ $10^{-5}$ to
$10^{-10}$ and search for the $e^{+}e^{-}$ or $\mu^{+}\mu^{-}$ decay of the 
dark photon via two signatures (invariant mass and displaced vertex). Using
Jefferson Lab's high luminosity electron beam along with a compact, large
acceptance forward spectrometer consisting of a silicon vertex tracker, lead
tungstate electromagnetic calorimeter and a muon detector, HPS will access
hitherto unexplored regions in the mass/coupling space. 

\end{Abstract}
\vfill
\begin{Presented}
DPF 2013\\
The Meeting of the American Physical Society\\
Division of Particles and Fields\\
Santa Cruz, California, August 13--17, 2013\\
\end{Presented}
\vfill
\end{titlepage}
\def\thefootnote{\fnsymbol{footnote}}
\setcounter{footnote}{0}

\section{The Physics Motivating the HPS Experiment}

The existence of additional $U(1)$ gauge symmetries of nature are ubiquitous in
several Beyond the Standard Model (BSM) theories 
\cite{Goodsell:2010ie,Candelas:1985en,Andreas:2011in,Jaeckel:2010ni}. Indeed,
it is natural for the associated gauge boson (heavy photon, dark photon or
$A'$) to ``kinematically mix'' with the Standard Model (SM) photon through the 
interaction of massive fields \cite{Holdom:1985ag}.  This, in turn, 
induces an effective coupling of the $A'$ to electric charge, which is 
suppressed relative to the electron charge by a factor of 
$\varepsilon \sim 10^{-2} - 10^{-12}$. 

The mixing of the photon with the $A'$ offers one of the few, 
non-gravitational, portals that can be used to search for new  
``hidden sector'' particles.  Some theoretical models have envisioned a 
scenario in which dark matter inhabits the hidden sector, with its interactions
mediated via an $A'$ 
\cite{ArkaniHamed:2008qn,Pospelov:2008jd,Cheung:2009qd,ArkaniHamed:2008qp}. 
Such models favor an $A'$ mass range on the order of 
$\sqrt{\varepsilon m_{W}} \sim$ MeV-GeV, in part because it resolves several
recently observed astrophysical phenomena \cite{Essig:2010ye}.  This includes
accounting for the observed excess in the cosmic-ray positron flux 
\cite{Adriani:2008zr, FermiLAT:2011ab, Aguilar:2013qda}, while remaining in 
accord with the lack of excess in the anti-proton spectrum 
\cite{Adriani:2010rc}. It must also be noted that an $A'$ with a mass in this
range could also provide an explanation for the anomalous magnetic moment of
the muon \cite{Pospelov:2008zw}.

\begin{figure}[t!]
\centering
\includegraphics[height=2.5in]{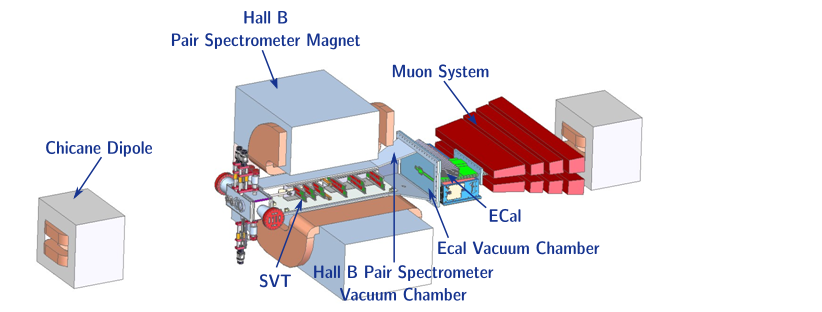}
\caption{The full HPS detector.  It will consist of a six layer silicon 
         microstrip tracker (SVT) installed inside the Hall B pair 
         spectrometer vacuum chamber.  A lead tungsten electromagnetic
         calorimeter (Ecal) will be installed downstream along with a 
         muon system.}
\label{fig:hps_detector}
\end{figure}

Sensitivity to this hitherto unexplored region of the mass-coupling phase 
space can be best achieved using high luminosity, fixed target experiments
\cite{Bjorken:2009mm}.  In such experiments, an electron beam incident on a 
high $Z$ target will produce dark photons through a process analogous to
bremsstrahlung, with the $A'$ subsequently decaying to pairs of fermions. Since
the electroproduced $A'$  will carry most of the incident electron beam energy,
the opening angle of low mass dark photons will be quite small.  Consequently,
the $A'$ decay products will be highly boosted, requiring a
detector with very forward acceptance that 
can be placed in close proximity to the target.  Maximizing
the acceptance will require placing the detector close to the beam plane, which
is occupied by the intense flux of multiple Coulomb scattered beam particles 
and radiative secondaries originating from the target.  This establishes a 
``dead zone'' that the detector is unable to encroach in order to avoid 
extensive radiation damage. Furthermore, the detector must also be operated in
vacuum in order to avoid additional background from beam gas interactions.  
Finally,  minimizing the material budget of the active area of the detector is
essential to reducing multiple scattering that dominates both the mass and 
vertex resolutions that determine the experimental sensitivity. These design 
principles have led to the conception of the HPS detector as shown in 
Fig. \ref{fig:hps_detector}.  The HPS detector will utilize a compact, large 
acceptance forward spectrometer consisting of a silicon microstrip tracker 
(SVT) along with a lead tungstate electromagnetic calorimeter (Ecal), used as 
the primary trigger, and a muon detector to search for dark photons in the mass
range of 20 MeV/$c^2$ to 1 GeV/$c^2$ and couplings in the range of 
$\epsilon \sim 10^{-2} - 10^{-5}$. What follows is a description of the HPS
experiment\footnote{The design of the muon system is a work in progress and 
will not be described here.} along with discussion of the results from the HPS
test run. 

\section{The HPS Detector}

The HPS experiment is proposed to run in experimental Hall B at Jefferson Lab 
using CEBAF's\footnote{Continuous Electron Beam Accelerator Facility} high 
luminosity electron beam. CEBAF's continuous duty cycle allows HPS to employ
detectors with short live times and rapid readouts in order to diminish 
backgrounds while maximizing the luminosity. The HPS setup will utilize a
three-magnet chicane system with the second dipole, the Hall B pair 
spectrometer magnet, serving as an analyzing magnet. The target will be placed
on the upstream edge of the analyzing magnet and will be 10 cm from the first
layer of the SVT, both of which will be housed within the Hall B pair 
spectrometer vacuum chamber. This setup has been designed to run with beam 
energies ranging from 1.1 GeV to 6.6 GeV and currents up to 500 nA incident on
a thin tungsten target of radiation lengths up to 0.25\% $X_{0}$. 

\begin{figure}[b!]
\centering
\includegraphics[height=2.5in]{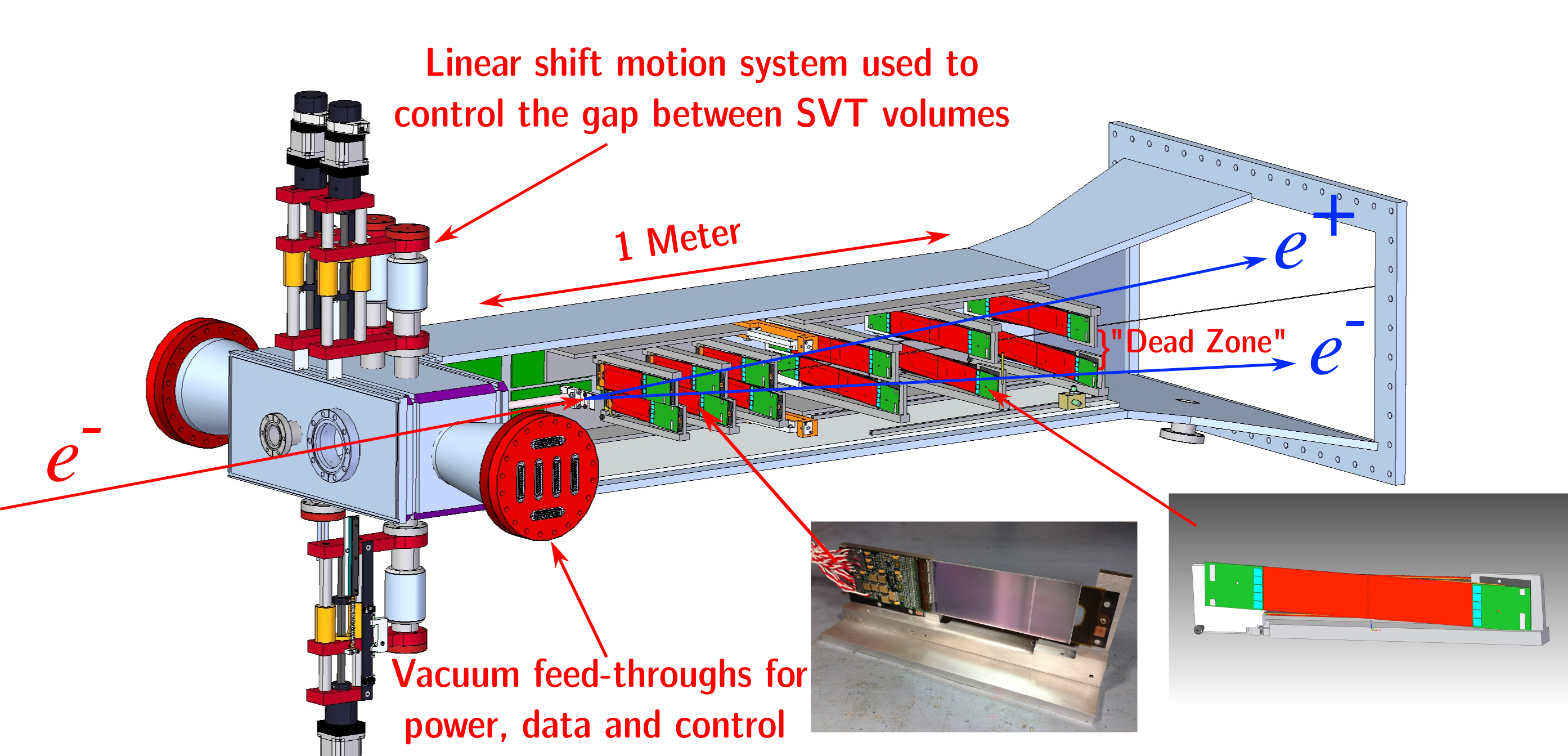}
\caption{A rendering of the HPS SVT showing the arrangement of the silicon 
		 planes inside of the Hall B pair spectrometer vacuum chamber. The 
		 silicon sensors are shown in red while the hybrid readout boards are 
		 shown in green.}
\label{fig:hps_svt}
\end{figure}

The SVT is comprised of six measurement layers, each consisting of a pair of 
closely-spaced silicon planes as shown in Fig. \ref{fig:hps_svt}.  A stereo
angle is introduced between the two planes within each layer allowing for the
measurement of both the vertical and bend coordinate of a hit, in turn, 
enabling full 3D hit reconstruction. The first three layers consist of 
Hamamatsu Photonics Corporation microstrip sensors and use a stereo angle of 
100 mrad. In order to better match the acceptance of the Ecal, the coverage of
the last three layers is two sensors wide and use a stereo angle of 50 mrad.
In total, the SVT will make use of 36 sensors, which amounts to 23,004 channels
(639 channels/sensor). The SVT layout is summarized in Table 
\ref{tab:svt_layout}.  

\begin{table}[b]
\begin{center}
\begin{tabular}{l|cccccc} 
\hline \hline 
Layer & 1 & 2 & 3 & 4 & 5 & 6 \\ \hline
$z$ position from target [cm]    & 10 & 20 & 30 & 50 & 70 & 90 \\
Stereo angle [mrad] & 100 & 100 & 100 & 50 & 50 & 50 \\
Non-bend plane resolution [$\mu$m] & $\approx6$ & $\approx6$ & $\approx6$
& $\approx6$ & $\approx6$ & $\approx6$ \\
Bend-plane resolution [$\mu$m] & $\approx60$ & $\approx60$ & $\approx60$
& $\approx120$ & $\approx120$ & $\approx120$ \\
\hline \hline
\end{tabular}
\caption{The proposed layout of the SVT.}
\label{tab:svt_layout}
\end{center}
\end{table}

The SVT sensors will be continuously read out at 40 MHz using the APV25 readout
chip operating in ``multi-peak'' mode with a 35 ns shaping time. In this mode of
operation, six samples of the shaped analog signal are read out per hit.  
Fitting these samples with a known CR-RC shape then allows for the determination
of the initial time of a hit with a resolution of $\approx$ 2 ns for a 
S/N $>$ 25. A S/N at this level should also result in a spatial resolution of 
$\approx6$ $\mu$m \cite{Azzi:1999}. 

The SVT is split into upper and lower tracking volumes in order to avoid the 15
mrad ``dead zone'', putting the active area of the sensors at 1.5 mm from the 
beam plane. The layers are mounted on upper and lower support structures that
are hinged on the downstream end of the SVT.  This allows adjustment of the
vertical position of the sensors remotely by a motion control system, in 
response to experimental conditions. 

\begin{figure}[t!]
\centering
\includegraphics[height=2.0in]{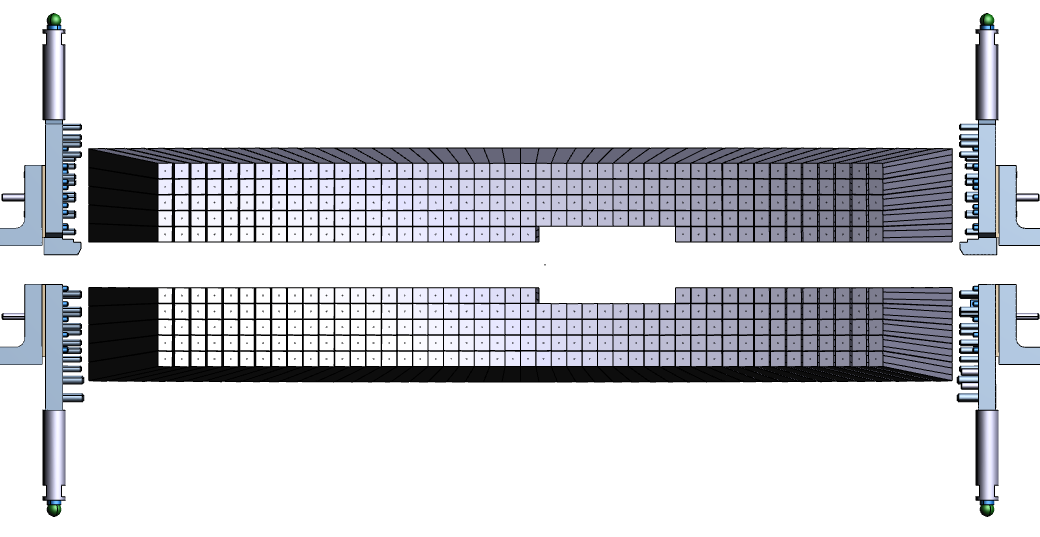}
\caption{A rendering showing the arrangement of the Ecal crystals. The Ecal is
		 split into upper and lower modules in order to accommodate the 
		 ``dead zone''.  The missing crystals allow a larger opening for the
		 outgoing electron beam.}
\label{fig:hps_ecal}
\end{figure}

The Ecal will be used as the primary trigger for the experiment as well as for
electron identification. It consists of two volumes (upper and lower), each 
consisting of five layers of PbWO$_{4}$ crystals arranged as shown in Fig. 
\ref{fig:hps_ecal}.  In total, this amounts
to 442 crystals. Each of the crystals are read out using avalanche photodiodes
at which point each of the output pulses are shaped and preamplified before 
being digitized by the JLAB FADC250, a 250 MHz flash ADC.  Pulse height 
information along with spatial and timing information is provided to the 
trigger by the FADC every 8 ns.
 

\section{The HPS Reach}

\begin{figure}[htb]
\centering
\includegraphics[height=3.5in]{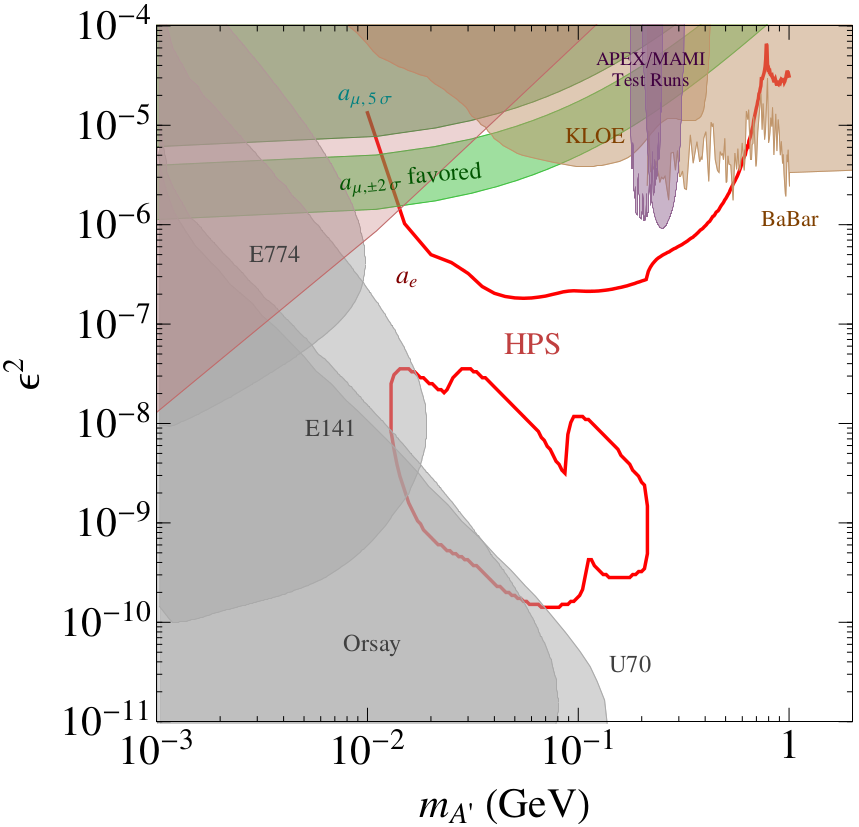}
\caption{The estimated reach of the HPS experiment at 2$\sigma$ significance 
		 along with existing constraints 
		 \cite{Pospelov:2008zw, Riordan:1987aw, Bross:1989mp, Andreas:2012mt, 
               Blumlein:2011mv, Aubert:2009cp, Essig:2010xa, Babusci:2012cr,
             Abrahamyan:2011gv, Merkel:2011ze, Endo:2012hp, Davoudiasl:2012ig}. 
         The reach calculation assumes running at 1.1 GeV
		 and 2.2 GeV for a week each in 2014 and running at
		 2.2 GeV and 6.6 GeV for three 
		 weeks each in 2015.}
\label{fig:hps_reach}
\end{figure}

The estimated reach of the HPS experiment at 2$\sigma$ significance is shown 
in Fig. \ref{fig:hps_reach} along with some existing constraints set by 
the beam dump experiments E141 \cite{Riordan:1987aw}, E774 \cite{Bross:1989mp},
Orsay \cite{Andreas:2012mt} and U70 \cite{Blumlein:2011mv}, the collider 
experiments BaBar \cite{Aubert:2009cp, Essig:2010xa} and KLOE 
\cite{Babusci:2012cr}, the fixed target test results reported by APEX 
\cite{Abrahamyan:2011gv}  and MAMI \cite{Merkel:2011ze} and the anomalous 
magnetic moment of the electron \cite{Endo:2012hp,Davoudiasl:2012ig} and muon
\cite{Pospelov:2008zw}. The green band represents the region that an $A'$
can be used to explain the discrepancy between the measured and calculated
muon anomalous magnetic moment.
The  reach calculation assumes running at 1.1 GeV and
2.2 GeV for a week each in 2014 as well as running at 2.2 GeV and 6.6 GeV for 
two weeks each in 2015. Sensitivity to the upper region is achieved through a
bump-hunt search while the lower region utilizes a bump-hunt plus displaced
vertex search.
 
\section{The HPS Test Run}

\begin{figure}[htb]
\centering
\includegraphics[height=3.5in]{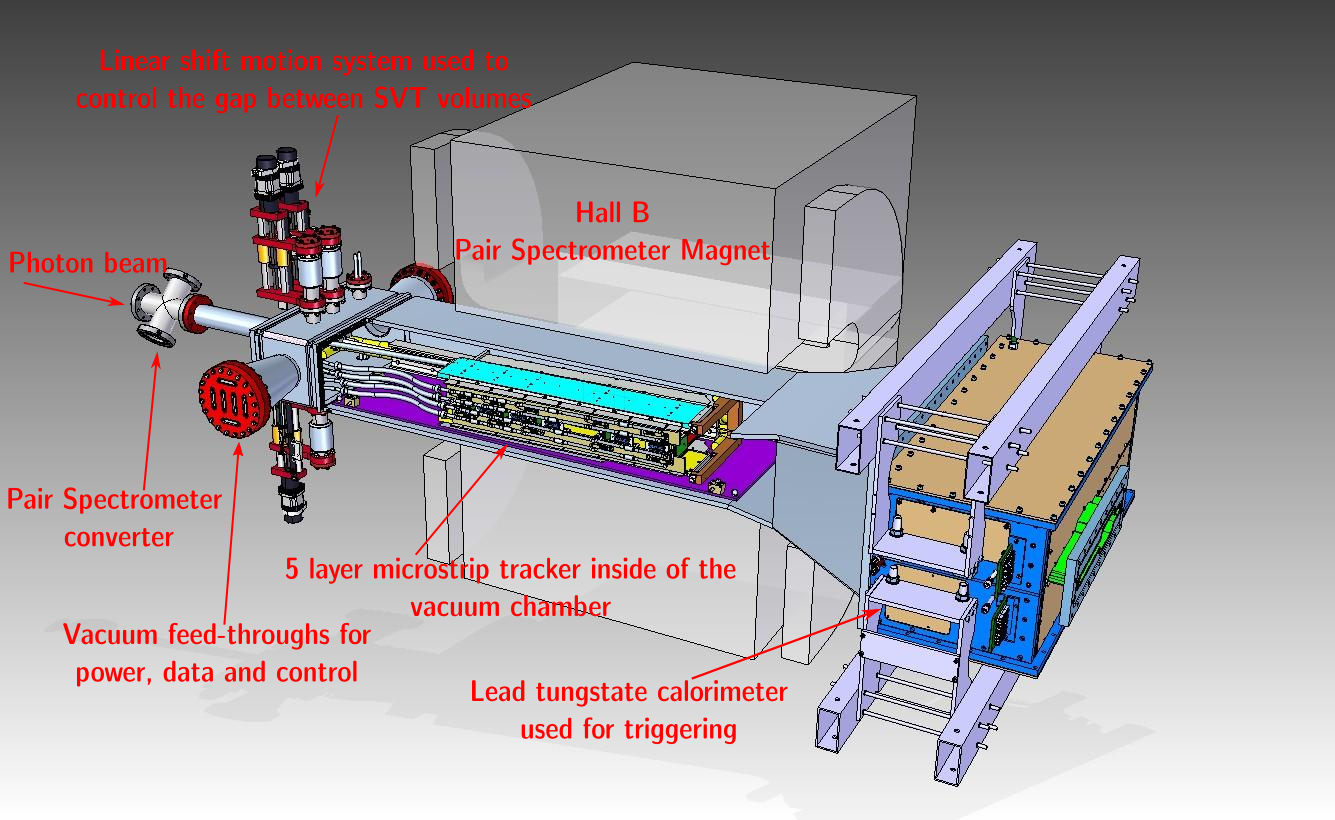}
\caption{The HPS test run detector.  It consist of a five layer silicon 
		 microstrip tracker installed inside of the Hall B pair spectrometer
		 vacuum chamber. A PbWO$_{4}$ electromagnetic calorimeter was installed
		 downstream and provided the trigger.  An aluminum converter upstream of the
		 tracker served as a target.}
\label{fig:hps_test}
\end{figure}

The HPS test run used a simplified version of the full HPS detector in order
to demonstrate the technical feasibility of the full HPS apparatus and to
confirm that the trigger rates and occupancies encountered during electron-beam
running are well understood.  The layout of the test run detector is shown in 
Fig. \ref{fig:hps_test}.  Other than the absence of the muon system, the
central difference to the design of the full detector was the smaller SVT 
acceptance due to the use of five layers with single sensor coverage. A more
thorough description of the test run detector geometry can be found in
\cite{Adrian:2013}.

The test run detector was designed with the intention of running using an 
electron beam. However, due to scheduling conflicts at JLab, dedicated electron
beam time was not secured. Instead, the experiment ran parasitically using
a photon beam with a thin aluminum converter upstream of the detector serving as a
target.  The conversions originating from the target were used to commission 
the detector and benchmark the performance of both the test SVT and Ecal.
HPS was also granted 8 hours of dedicated photon beam before the CEBAF 
shutdown.  Several runs at currents ranging from 30-90 nA using different 
converter thicknesses of radiation lengths 1.6\%, 0.45\% and 0.18\% $X_{0}$
were taken. The dedicated run was used to measure the normalized trigger rates
that established that the backgrounds expected during electron running are 
well modeled.  

Throughout the duration of the test run, $>$ 97\% of the SVT's channels were
found to be operational with a signal to noise ratio of $\approx$ 25.5.  This
is enough to achieve a position resolution of $\approx$ 6 $\mu$m desired for the
full HPS run. The hit time resolution was found to be $\approx$ 2.6 ns, which is
enough to aid in the reduction of background due to pileup.  The single hit
efficiency was also shown to exceed 98\% for good channels while tracks were
reconstructed with very high efficiency and purity.  Finally, the survey-based
alignment of the SVT was shown to be adequate enough to allow the use of 
track-based alignment. 

The Ecal performed similarly well with 87\% of the crystals found to be 
operational. The trigger functioned as designed and tests showed that it can 
achieve trigger rates greater than 100 kHz.

The requirement that the SVT be as close to the beam plane as possible creates
one of the key challenges for the HPS experiment. Simulations have shown that
the dominant source of occupancy for the HPS experiment comes from beam 
electrons that have multiple Coulomb scattered in the target into large angles.
However, the multiple 
Coulomb scattering rate at large angles is overestimated by Geant4 by a factor
of two when compared to EGS5 \cite{Adrian:2013}.  It 
is then crucial to verify that the correct model of multiple Coulomb scattering
is being used by the simulation in order to establish that HPS can run at the
proposed electron beam currents.

\begin{figure}[htb]
\centering
\includegraphics[height=2.0in]{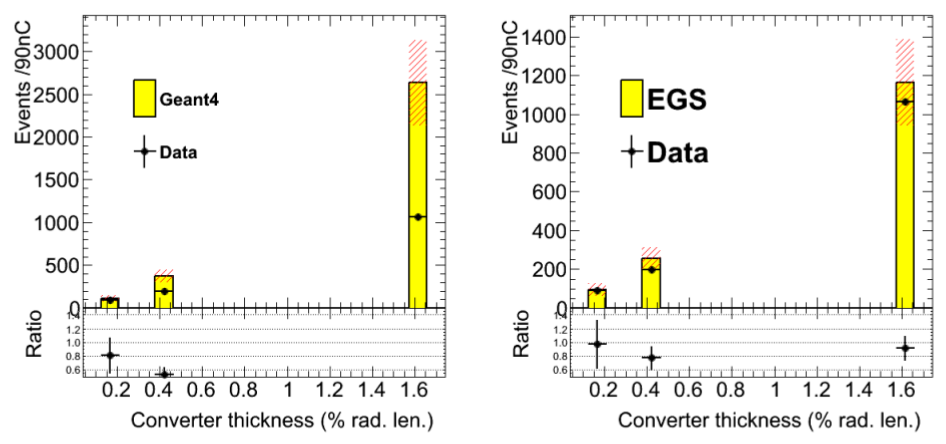}
\caption{The normalized trigger rates as a function of converter thickness. The
		 trigger rates simulated by EGS5 agree with the test run data to within
		 10\%.}
\label{fig:rates}
\end{figure}

The angular distribution of the conversions produced in the aluminum target are a 
convolution of both the pair production angle and the multiple Coulomb 
scattering of the electron-positron pairs.  It is then possible to confirm the 
model of multiple Coulomb scattering by using the data taken during the 
dedicated photon run. Comparing the measured angular distribution for each of
the converter thicknesses to those from simulation verifies that EGS5 predicts
the correct distributions while Geant4 overestimates the rates. In fact, the 
normalized trigger rates as simulated by EGS5 agree with the test run data to 
within 10\% as shown in Fig. \ref{fig:rates}.  This verifies the EGS5 model of 
multiple Coulomb scattering and provides confidence that the beam backgrounds 
expected during the full run are as simulated. 
 
\section{Conclusion}

The HPS test run successfully demonstrated that the design of the full detector
is technically feasible.  Specifically, the SVT was able to demonstrate that 
the signal to noise, timing and position resolutions were as expected while the 
Ecal trigger functioned as designed.  In addition, the EGS5 model of multiple 
Coulomb was verified. After the test run, a proposal for the full HPS 
experiment was submitted to the Department of Energy and has now been approved.
Construction of the full experiment is now underway and should be ready for
installation and data taking in the Fall of 2014.

\end{document}

%% file: econfmacros.tex
\def\beq{\begin{equation}}
\def\eeq#1{\label{#1}\end{equation}}
\def\eeqn{\end{equation}}

\def\beqa{\begin{eqnarray}}
\def\eeqa#1{\label{#1}\end{eqnarray}}
\def\eeqan{\end{eqnarray}}

\let\bar=\overbar

\def\Dslash{\not{\hbox{\kern-4pt $D$}}}
\def\dslash{\not{\hbox{\kern-2pt $\del$}}}

\def\msb{{\bar{\ssstyle M \kern -1pt S}}}

